\documentclass[pre,twocolumn]{revtex4}
\usepackage[english]{babel}
\usepackage{amsmath}
\usepackage{amssymb}
\usepackage{epsfig}

\begin{document}

\title{Waves over Curved Bottom: The Method of Composite Conformal Mapping}
\author{Victor P. Ruban}
\email{ruban@itp.ac.ru}
\affiliation{Landau Institute for Theoretical Physics, RAS,
Chernogolovka, Moscow region, 142432 Russia} 
\date{\today}

\begin{abstract}
A compact and efficient numerical method is described 
for studying plane flows of an ideal fluid
with a smooth free boundary over a curved and nonuniformly
moving bottom. Exact equations of motion in
terms of the so-called conformal variables are used. 
In addition to the previously known applications for shear
flows with constant (including zero) vorticity, here 
a generalization is made to the case of potential flows in
uniformly rotating coordinate systems, where centrifugal 
and Coriolis forces are added to the gravity force. A
brief review is given of previous results obtained by 
this method in a number of physically interesting problems
such as modeling of tsunami waves caused by the movement 
of nonuniform bottom, the dynamics of Bragg
(gap) solitons over a spatially periodic bottom profile, 
the Fermi-Pasta-Ulam (FPU) recurrence phenomenon 
for waves in a finite pool, the formation of anomalous 
waves in an opposing nonuniform current, and the
propagation of a solitary wave in a shear current and 
its runup on a depth difference. In addition, a number
of new numerical results are presented concerning the 
nonlinear dynamics of a free boundary in closed rotating 
containers partially filled with a fluid -- centrifuges 
of complex shape. In this case, the equations of motion
differ in some essential details from those of $x$-periodic systems.
\end{abstract}

\maketitle

\section{Introduction}

In some branches of hydrodynamics, the approximation 
of an ideal incompressible fluid turns out to be
quite acceptable. In particular, this includes the theory
of surface waves on water. The description of flows is
significantly simplified if we assume that they are
potential. Moreover, since the Laplace equation for
the velocity field potential in the spatially two-dimensional
case is conformally invariant, and we are talking
about domains with curved boundaries, the theory of
conformal mappings and the corresponding analytic
functions is suitable for the occasion. Therefore, it is
not surprising that the idea of using conformal mappings
for the exact representation of the equations of
nonlinear dynamics of potential surface waves arose
quite a long time ago (in Ovsyannikov's works [1, 2]).
For a long time, these equations remained unclaimed,
and few people knew about them. Only with the development 
and extensive spread of computer technology
in the 1990s and with the effective practical implementation 
of fast discrete Fourier transform algorithms, 
the equations of motion of a free surface in
conformal variables were rediscovered by Zakharov
and his colleagues [3-6] and became a working tool
for many researchers (see [7-34] and numerous references 
therein). The authors mainly modeled waves on
infinitely deep water, for which the most important
results had been obtained over two decades, including
the pioneering observations of anomalous waves (the
so-called rogue waves) in high-precision numerical
experiments of Zakharov's group [7-10]. However,
there are a number of interesting problems in which
the effects of the interaction of waves with a nonuniform 
bottom profile are fundamentally important. To
study these problems, in 2004 the present author generalized 
the method of conformal variables to the case
of a fixed curved bottom [35]. The waves over a moving 
bottom were examined in [36], and shear flows
with a free boundary over a variable depth were modeled 
by the conformal mapping method in [37]. The
dynamic objects in this description are the potential
component of the velocity field and the conformal
mapping of the horizontal strip $0\leq v\leq 1$ from an auxiliary 
complex plane $w=u+iv$ to the moving domain
in the vertical plane $z=x+iy$ occupied by the fluid.
The conformal mapping $z(w,t)=Z(\zeta(w,t),t)$ can be
represented as a composition of two analytic functions:
an unknown function $\zeta(w,t)$ and the given function 
$Z(\zeta,t)$; the first of these functions leaves invariant
the real axis (i.e., $\mbox{Im}\,\zeta(u,t)=0$), and the second 
parametrically defines the shape of the bottom 
$X^{(b)}(s,t)+iY^{(b)}(s,t)=Z(s,t)$.
As a result, a single unknown real function 
$a(u,t)=\mbox{Re}\,\zeta(u+i,t)$ is sufficient to parameterize 
the shape of the free boundary. The complex
velocity potential with regard to the kinematic condition
at the lower boundary also depends on a single
unknown real function $\psi(u,t)$. The generalized Bernoulli 
equation and the kinematic condition on the
upper (free) boundary imply the equations of motion
for $a(u,t)$ and $\psi(u,t)$. The equations contain linear
operators that are diagonal in the Fourier representation, 
so that a pseudospectral numerical method using
fast Fourier transform turns out to be the most natural
and convenient.

Since then, a number of results have been obtained
that confirmed the practical advantage and high efficiency 
of the method. In particular, the dynamics of
the Bragg (gap) solitons over a spatially periodic bottom 
profile [38,39], the Fermi-Pasta-Ulam (FPU)
recurrence phenomenon for waves in a finite pool [40,41], 
and the formation of anomalous waves on an
opposing nonuniform current [42] have been investigated. 
The goal of the present study is to give a brief
overview on this topic and present some new results on
the classical problem of the free surface dynamics in
partially filled closed rotating containers, which has
not been previously discussed in connection with conformal 
variables (see, for example, [43-45] and references 
therein). In fact, one can talk about quasi-two-dimensional
centrifuges (irregularly shaped cylinders
with characteristic transverse dimension $R$ and length
$h\ll R$, or eccentric circular cylinders) that rotate
around a horizontal axis (the axis is parallel to the generatrices) 
and may be subjected to angular and/or linear
accelerations, as well as to deformations of the
cross-sectional shape. At sufficiently high rotational
speeds of $\Omega\gg \nu/h^2$ (where $\nu$ is the kinematic viscosity), 
viscous effects become insignificant at times on
the order of several tens of $\Omega^{-1}$, which corresponds to
many revolutions. It is very important that the fluid
flow is not potential at all, but is close to solid state
rotation with angular velocity $\Omega$. Therefore, a natural
step is the transition to a rotating coordinate system in
which, as is known, the centrifugal and Coriolis forces
are added to the gravity force. Perturbations of the
(divergence-free) velocity field due to the accelerations
and deformations of the container turn out to be
purely potential, so that a generalized Bernoulli equation 
holds, in which the stream function harmonically
conjugate to the potential appears as a separate term
(due to the Coriolis force). It should be noted that
even a relatively simple problem of nonlinear waves in
a coaxial circular container in the absence of the gravity 
force has been little studied, although it is a very
elegant and informative problem, not to mention the
full version with deformable noncircular containers
and their accelerations. Therefore, the use of conformal 
variables looks very promising here.

Further presentation is organized as follows. In
Section 2, we give a fairly detailed description of the
method used, including the derivation of the exact
equations of motion and the discussion of the numerical 
scheme. In Section 3, we briefly discuss the most
interesting previous results obtained by the author
using a composite conformal mapping. Section 4 is
devoted to a new application of conformal variables to
describe surface waves in rotating systems. Finally,
Section 5 contains a brief conclusion.

\section{Description of the method}
\subsection{Exact equations of motion}

We will derive exact and explicit equations of
motion of a free boundary for the case of waves in centrifuges, 
since systems infinite along $x$ are largely similar 
and have been considered in detail in the above-cited papers [35-37].

As is known, the Euler equation for plane flows of
an ideal incompressible fluid in a rotating (counterclockwise) 
coordinate system contains additional forces (componentwise)
$(\Omega^2x,\Omega^2y)+(2\Omega V^{(y)},-2\Omega V^{(x)})$
and admits the reduction 
${\bf V}=(V^{(x)},V^{(y)})=(\varphi_x, \varphi_y)=(\theta_y,-\theta_x)$, 
where the subscripts denote partial derivatives, the function 
$\varphi(x,y,t)$ is the potential of
the velocity field, and $\theta(x,y,t)$ is the corresponding
harmonically conjugate stream function. Therefore,
the generalized Euler equation has the form
\begin{eqnarray}
\varphi_t+(\varphi_x^2+\varphi_y^2)/2-\Omega^2(x^2+y^2)/2
+2\Omega\theta&&\nonumber\\
+g(t)(y\cos\Omega t+x\sin\Omega t)+\tilde P=0,&&
\label{Euler_generalized}
\end{eqnarray}
where $\tilde P$ is pressure divided by the density of the fluid.
The effective gravity field $g(t)$ depends on time if the
axis of rotation is subjected to vertical accelerations.

Suppose that, at each instant of time, the flow
domain represents a deformed disk with one ``hole'' ---
a free surface. Consider a scalar function $v(x,y,t)$ satisfying 
the Laplace equation  $v_{xx}+v_{yy}=0$ and taking
fixed boundary values: $v=0$ on the outer boundary of
the domain (i.e., at the bottom of the container), and
$v=1$ on the free (inner) boundary. Such a function
exists and is unique. Now, let us construct a function
$u(x,y,t)$ harmonically conjugate to $v$. This function is
multivalued (and is defined up to an additive constant).
Its increment $L(t)$ when going along the free
surface (the conformal modulus) generally depends on
time. We compose a complex combination $w=u+iv$,
which is an analytic function of the complex variable
$z=x+iy$. It is clear that such a curvilinear change of
coordinates corresponds to a conformal mapping $z(w,t)$
of a rectangle with dimensions $L\times 1$; moreover,
the variable $v$ is an analog of the radial coordinate, and
the variable $u$ is an analog of the angular coordinate.

Now the shape of the free boundary is defined
parametrically by the formula
\begin{equation}\label{surface_shape}
X^{(s)}(u,t)+iY^{(s)}(u,t)\equiv Z^{(s)}(u,t)=z(u+i,t),
\end{equation}
and the bottom profile, by the formula
\begin{equation}\label{bottom_shape}
X^{(b)}(u,t)+iY^{(b)}(u,t)\equiv  Z^{(b)}(u,t)=z(u,t).
\end{equation}

It is very important that the complex potential 
$\varphi(u,v,t)+i\theta(u,v,t)=\phi(w,t)$ is an analytic function.
Denote the boundary values of this function as follows:
\begin{equation}\label{boundary_potential}
\phi(u+i,t)\equiv \Phi^{(s)}(u,t),\qquad \phi(u,t)\equiv \Phi^{(b)}(u,t).
\end{equation}
Since $\Phi^{(s)}(u,t)$ and $\Phi^{(b)}(u,t)$ are values of the same
analytic function at the points $u$ and $u+i$, they are
related by a linear transformation (see [36]):
\begin{equation}\label{Phi_s_Phi_b_relation}
\Phi^{(s)}(u,t)=e^{-\hat k}\Phi^{(b)}(u,t),
\end{equation}
where $e^{-\hat k}\equiv\exp(i\hat\partial_u)$. This implies the relation
$\Phi^{(s)}_k(t)=e^{-k}\Phi^{(b)}_k(t)$ for the corresponding Fourier
transforms. For further exposition, we need a few
more linear operators. These operators $\hat S$, $\hat R$,
and $\hat T=\hat R^{-1}$ are also diagonal in the Fourier representation:
\begin{equation}\label{SR}
 S_k={1}/{\cosh(k)},\quad R_k=i\tanh(k),\quad  T_k=-i\coth(k).
\end{equation}
It is important that the action of any of these operators
on a real function yields a purely real function.

Equation (1), rewritten in new variables, has the
following form on the free boundary (the dynamic boundary condition):
\begin{eqnarray}
&&\mbox{Re}\left(\Phi^{(s)}_t-
\Phi^{(s)}_u{Z^{(s)}_t}/{Z^{(s)}_u}\right) 
+|\Phi^{(s)}_u/Z^{(s)}_u|^2/2\nonumber\\
\label{Bernoulli_Dw}
&&\quad-\Omega^2|Z^{(s)}|^2/2 +2\,\Omega\,\mbox{Im\,}\Phi^{(s)}\nonumber\\
&&\quad+g(t)\mbox{Im\,}[Z^{(s)}\exp(i\Omega t)]=0.
\label{Euler_generalized_u}
\end{eqnarray}
To this equation, we add two kinematic boundary conditions:
\begin{equation}\label{kinematic_Dw_s}
\mbox{Im}\big(Z^{(s)}_t\bar Z^{(s)}_u\big)=-\mbox{Im\,}\Phi^{(s)}_u,
\end{equation}
\begin{equation}\label{kinematic_Dw_b}
\mbox{Im}\big(Z^{(b)}_t\bar Z^{(b)}_u\big)=-\mbox{Im\,}\Phi^{(b)}_u,
\end{equation}
where $\bar Z$  denotes the complex conjugate. It is convenient 
to represent $\Phi^{(b)}(u,t)$ as
\begin{equation}\label{Phi_b}
\Phi^{(b)}(u,t)=\hat S \psi(u,t) -i (1-i\hat R) \hat\partial_u^{-1}f(u,t),
\end{equation}
where $\psi(u,t)$ and $f(u,t)$ are some unknown real functions. 
Then, due to relation (5) and the operator equalities
\begin{equation}
e^{-\hat k}\hat S=(1+i\hat R),\qquad e^{-\hat k}(1-i\hat R)=\hat S, 
\end{equation}
we obtain the following formula for $\Phi^{(s)}(u,t)$:
\begin{equation}\label{Phi_s}
\Phi^{(s)}(u,t)=(1+i\hat R)\psi(u,t) -i \hat S  \hat\partial_u^{-1}f(u,t).
\end{equation}
Next, we use the following function in the equations:
\begin{equation}\label{Psi_def}
\Psi(u,t)\equiv (1+i\hat R)\psi(u,t).
\end{equation}

Now we should take into account that the function 
$z(w,t)$ can be represented as a composition of two
functions (see [35,36]), that is, $z(w,t)=Z(\zeta(w,t),t)$,
where the known analytic function $Z(\zeta,t)=X(\zeta,t)+iY(\zeta,t)$
defines the shape of the bottom by the mapping 
of the real axis. The conformal mapping $Z(\zeta,t)$
should not have singularities within a sufficiently wide
horizontal strip over the real axis in the plane $\zeta$, so that
there is ``enough space'' for large-amplitude waves in
the $z$ plane (in a particular case, $Z(\zeta,t)$ may not have
singularities in the entire upper half-plane $\zeta$). The
decomposition of the conformal mapping $z(w,t)$ into a
composition is not unique, because if $Z(\zeta,t)=Z_1(\beta(\zeta,t),t)$,
where the analytic function $\beta(\zeta,t)$ leaves
invariant the real axis and does not have singularities
too close to it, then

$$Z(\zeta(w,t),t)=Z_1(\beta(\zeta(w,t),t),t)=Z_1(\zeta_1(w,t),t).$$

The intermediate analytic function $\zeta(w,t)$ takes
real values on the real axis, and therefore the equality
\begin{equation}\label{zeta_def}
\zeta(w,t)=\int \frac{a_k(t)}{\cosh(k)}e^{ikw}\frac{dk}{2\pi}, 
\qquad a_{-k}=\bar a_k,
\end{equation}
is valid, where $a_k(t)$ is the Fourier transform of some
real function $a(u,t)$. At the bottom, $\zeta(u,t)=\hat S a(u,t)$,
and therefore
\begin{equation} \label{Zb}
Z^{(b)}(u,t)=Z(\hat S a(u,t),t), 
\end{equation}
On the free surface, we have the relations
\begin{equation}\label{xi_def}
\zeta(u+i,t)\equiv\xi(u,t)=(1+i\hat R)a(u,t),
\end{equation}
\begin{equation}
Z^{(s)}=Z(\xi,t),\quad Z^{(s)}_u=Z_\xi(\xi,t)\xi_u, 
\label{Z_Zu}
\end{equation}
\begin{equation}
Z^{(s)}_t=Z_\xi(\xi,t)\xi_t+Z_t(\xi,t).
\label{Zt}
\end{equation}

Since our goal is to derive equations of motion for
the unknown functions $\psi(u,t)$, $f(u,t)$ and $\xi(u,t)$, we
should substitute expressions (10), (12), (15), (16),
(17), and (18) into Eqs. (7), (8), and (9). Equation (9)
does not require any effort and immediately gives an
explicit relation
\begin{equation}
f=\mbox{Im\,}
\Big(Z_t(s,t)\bar Z_s(s,t)\hat S a_u\Big)\big|_{s=\hat S a},\quad
a=\mbox{Re\,}\xi.
\label{f}
\end{equation}
Equation (8), divided by $|Z^{(s)}_u|^2$, takes the form
\begin{equation}\label{kinematic_Dw_s_2}
\mbox{Im\,}\left(\frac{\xi_t}{\xi_u}\right)+
\mbox{Im\,}\left(\!\frac{Z_t(\xi,t)}{Z_\xi(\xi,t)\xi_u}\!\right)
=\frac{-\mbox{Im\,}\Psi_u +\hat S f}{|Z_\xi(\xi,t)\xi_u|^2}.
\end{equation}
Thus, we have $\mbox{Im}({\xi_t}/{\xi_u})=-Q$, where
\begin{equation}\label{Q_def}
Q\equiv
\frac{\mbox{Im\,}\Psi_u -\hat S f}
{|Z_\xi(\xi,t)\xi_u|^2}+\mbox{Im}\left(
\frac{Z_t(\xi,t)}{Z_\xi(\xi,t)\xi_u}\right).
\end{equation}
Next, we argue as follows: since
$$\xi_t/\xi_u=\zeta_t(w,t)/\zeta_w(w,t)|_{w=u+i}$$ 
is a value of the analytic function taken at $u+i$, that is
real on the real axis, there is the following relation
between the real and imaginary parts: 
$\mbox{Im}(\xi_t/\xi_u)=\hat R\,\mbox{Re}(\xi_t/\xi_u)$, 
so that $\mbox{Im}({\xi_t}/{\xi_u})=-Q$ implies 
$\xi_t=-\xi_u(\hat T+i)Q$. This yields an equation for $a_t$:
\begin{equation}\label{a_t} 
a_t=-\mbox{Re}[\xi_u(\hat T+i)Q].
\end{equation}
Now we substitute the necessary expressions into
Eq. (7) in order to find $\psi_t$. Straightforward transformations 
lead to the following equation:
\begin{eqnarray}
\psi_t&&=-\mbox{Re}[\Psi_u\,(\hat T+i)Q]
+\mbox{Re}\left(\frac{\Psi_u Z_t(\xi,t)}{Z_\xi(\xi,t)\xi_u}\right)\nonumber\\
&&-\frac{|\Psi_u|^2 - (\hat S f)^2}{2|Z_\xi(\xi,t)\xi_u|^2}
-g(t)\mbox{Im}[Z(\xi,t)\exp(i\Omega t)]\nonumber\\
&&+\frac{\Omega^2}{2}|Z(\xi,t)|^2 
-2\Omega(\mbox{Im}\Psi-\hat S\hat\partial_u^{-1} f).
\label{psi_t}
\end{eqnarray}
Thus, we have derived exact equations of motion for
waves on the free surface of an ideal fluid in centrifuges. 
As for shear flows in systems infinite along $x$,
everything is done similarly, except that a different
generalized Bernoulli equation is used (see [37]).

\subsection{Numerical simulation}

For the practical application of system (19), (21),
(22), and (23) in numerical simulation, one should
take into account two important facts discussed below.

First, the period $L$ in the variable $u$ (the previously
mentioned conformal modulus of the mapping)
depends on time due to the singularity of the operator
$\hat T\approx 1/(ik)=\hat\partial_u^{-1}$
for small $k$, which gives rise to a component $-a_u \langle Q\rangle u$, 
aperiodic in $u$, on the right-hand side
of Eq. (22) and a similar component $-\psi_u \langle Q\rangle u $ on the
right-hand side of Eq. (23), where $\langle Q\rangle$ is the average
value of $Q$  over the period $L$. Therefore, instead of the variable
$u$, it is more convenient to use the new variable $\vartheta=[2\pi/L(t)]u$,
for which the period is fixed and
equal to  $2\pi$. In this case, the equation of motion for
$\alpha(t)=2\pi/L(t)$ is obtained from the requirement that
the terms $\dot\alpha u a_\vartheta$ and $\dot\alpha u \psi_\vartheta$, 
aperiodic in $\vartheta$, on the left-hand 
sides of Eqs. (22) and (23) should cancel out with
the corresponding aperiodic components on the right-hand 
sides when substituted into these equations.
Obviously, the cancellation occurs for $\dot\alpha=-\alpha\langle Q\rangle$.

Second, if the period-average value of $f$ is different
from zero (which inevitably occurs when the walls of
the centrifuge are deformed with a change in its cross-sectional 
area ${\cal A}_c$), then an additional potential circular 
flow with circulation $\Gamma(t)$ appears in the system, so
that $\langle\psi_\vartheta\rangle=\Gamma/2\pi$, 
and the potential itself turns out to
be a multivalued function. It is useful to keep in mind
that, due to the conservation of the velocity circulation
along the free boundary in the laboratory coordinate
system, the integral of motion $\Gamma(t)+2\Omega{\cal A}_c(t)=const$
holds. This conservation law must be checked when
debugging the numerical code.

Thus, we can single out the fixed aperiodic part in
$a(\vartheta,t)$ by writing  $a(\vartheta,t)=\vartheta+\rho(\vartheta,t)$, 
where $\rho(\vartheta,t)$ is a $2\pi$-periodic function:
\begin{equation}
\rho(\vartheta,t)=\sum_{m=-\infty}^{+\infty}\rho_m(t)\exp(im\vartheta). 
\label{a_m} 
\end{equation}
Similarly, for the derivative $\psi_\vartheta$, we have
\begin{equation}
\psi_\vartheta(\vartheta,t) =\sum_{m=-\infty}^{+\infty}D_m(t)\exp(im\vartheta),
\label{Dpsi_m}
\end{equation}
where $D_0=\Gamma/2\pi$, $\rho_{-m}=\bar \rho_{-m}$, and $D_{-m}=\bar D_{-m}$. 
In this case, the formulas for the corresponding analytic
functions on the free boundary have the form
\begin{equation}
\xi(\vartheta,t)=\vartheta+i\alpha(t)
+\sum_{m=-\infty}^{+\infty}\frac{2\rho_m(t)\exp(im\vartheta)}{1+\exp(2m\alpha(t))}, 
\label{xi_m} 
\end{equation}
\begin{equation}
\Psi_\vartheta(\vartheta,t) =
\sum_{m=-\infty}^{+\infty}\frac{2D_m(t)\exp(im\vartheta)}{1+\exp(2m\alpha(t))}.
\label{DPsi_m}
\end{equation}
In view of the above remarks, the equations of motion
for the functions $\rho$ and $\psi_\vartheta$ in the variables 
$(\vartheta,t)$ look
similar to Eqs. (22) and (23), except that all $u$-derivatives 
should be replaced by $\vartheta$-derivatives and the previous 
operators $\hat R$, $\hat S$, and $\hat T$ should be replaced everywhere 
by new operators $\hat{\mathsf R}_\alpha$, 
$\hat{\mathsf S}_\alpha$, and $\hat{\mathsf T}_\alpha$, respectively:
\begin{eqnarray}
\rho_t&=&-\mbox{Re}[\xi_\vartheta(\hat {\mathsf T}_\alpha+i){\mathsf Q}],
\label{rho_t_alpha} 
\\
\psi_{\vartheta t}&=&\frac{\partial}{\partial\vartheta}\Bigg\{
\mbox{Re}\Big(\frac{\Psi_\vartheta Z_t(\xi,t)}{Z_\xi(\xi,t)\xi_\vartheta}\Big)
-\mbox{Re}[\Psi_\vartheta(\hat{\mathsf T}_\alpha +i){\mathsf Q}] 
\nonumber\\
&&-\frac{|\Psi_\vartheta|^2 - (\hat {\mathsf S} F)^2}{2|Z_\xi(\xi,t)\xi_\vartheta|^2}
-g(t)\mbox{Im}[Z(\xi,t)\exp(i\Omega t)]\nonumber\\
&&+\frac{\Omega^2}{2}|Z(\xi,t)|^2\Bigg\} 
-2\Omega(\mbox{Im}\Psi_\vartheta-\hat {\mathsf S} F),
\label{psi_t_alpha}
\end{eqnarray}
where $\xi=\vartheta+i\alpha+(1+i\hat{\mathsf R}_\alpha)\rho$, 
$\Psi_\vartheta=(1+i\hat{\mathsf R}_\alpha)\psi_\vartheta$,
\begin{eqnarray}
F&=&\mbox{Im\,}\Big(Z_t(s,t)\bar Z_s(s,t) 
(1+\hat {\mathsf S}\rho_\vartheta)\Big)\big|_{s=\vartheta+\hat {\mathsf S} \rho},\\
{\mathsf Q}&=& \frac{\mbox{Im\,}\Psi_\vartheta -\hat {\mathsf S} F}
{|Z_\xi(\xi,t)\xi_\vartheta|^2}+\mbox{Im}\left(
\frac{Z_t(\xi,t)}{Z_\xi(\xi,t)\xi_\vartheta}\right).
\end{eqnarray}
These new operators are diagonal in the discrete Fourier representation:
${\mathsf R}_\alpha(m)=i\tanh(\alpha m)$,
${\mathsf S}_\alpha(m)=1/\cosh(\alpha m)$, and
${\mathsf T}_\alpha(m)=-i\coth(\alpha m)$ for $m\not=0$, while
${\mathsf T}_\alpha(0)=0$.
Note that the operator $\hat{\mathsf T}_\alpha$ 
has no singularity. The system is closed by the following 
condition for $\dot\alpha(t)$, which guarantees the cancellation 
of aperiodic terms in Eqs. (22) and (23):
\begin{equation}\label{dot_alpha}
\dot\alpha(t)=-\frac{1}{2\pi}\int_0^{2\pi}{\mathsf Q}(\vartheta)d\vartheta.
\end{equation}

This system of equations has an integral of motion
corresponding to the conservation of the area occupied 
by the fluid (the centrifuge area ${\cal A}_c$ minus the
area of the hollow domain ${\cal A}_h$),
\begin{eqnarray}
{\cal A}&=&\frac{1}{2}\oint(X^{(b)} d Y^{(b)}- Y^{(b)} d X^{(b)})\nonumber\\
&-&\frac{1}{2}\oint(X^{(s)} d Y^{(s)}- Y^{(s)} d X^{(s)}).
\end{eqnarray}
In addition, if the container rotates uniformly at a certain 
angular velocity $\Omega+\Delta$ without changing its shape,
i.e., $Z(\zeta,t)=\exp(i\Delta t){\cal Z}(\zeta)$, then, for $g=0$, the sum of
the kinetic and centrifugal energies in the corresponding 
coordinate system is conserved (the fluid vorticity
in the laboratory system is still assumed to be equal
to $2\Omega$). If we denote $X+iY=Z^{(s)}\exp(-i\Delta t)$ and
$X_b+iY_b=Z^{(b)}\exp(-i\Delta t)$, then the conserved energy is
\begin{eqnarray}
{\cal E}_\Delta&=&\frac{1}{8}(\Omega^2+2\Omega\Delta)\oint (X^2+Y^2)(XdY-YdX)\nonumber\\
&+&\frac{\Delta^2}{8}\oint (X_b^2+Y_b^2)\hat{\mathsf R}_\alpha(X_b^2+Y_b^2)_\vartheta\,
d\vartheta\nonumber\\
&+&\frac{\Delta}{2}\oint[X^2+Y^2]\psi_\vartheta\,d\vartheta
-\frac{\Delta}{2}\oint[X_b^2+Y_b^2]\hat{\mathsf S}\psi_\vartheta\,d\vartheta\nonumber\\
&\!-\!&\frac{1}{2}\!\oint\!(\psi-\Gamma \vartheta/2\pi)\,\hat{\mathsf R}_\alpha 
\psi_\vartheta\,d\vartheta +\Gamma^2\alpha/4\pi.
\end{eqnarray}
If $\Delta=-\Omega$ and $g=const$ (the container is fixed in the
laboratory coordinate system), the sum of the kinetic
and potential energies in the gravity field is conserved:
${\cal E}_{-\Omega}+(g/2)\oint Y^2dX=const$.
 When testing a computer program, all of these conservation 
laws should also be checked. Otherwise, as
the computational practice has shown, one can make
an error that does not show up in all cases and may
therefore remain unnoticed.

Equations (28)-(32) are very convenient and easy
for numerical simulation if the function $Z(\zeta,t)$ is given
by a sufficiently compact explicit formula, as is the
case for many practically interesting bottom profiles.
In addition, the C programming language has data
type {\tt complex}, and the mathematical library contains
elementary functions of a complex variable, such as
{\tt cexp, clog}, etc., which makes the process of writing
a code quite simple and pleasant. The numerical
scheme created by the author is naturally based on the
discrete Fourier transform, since all linear operators in
the equations can be effectively calculated in the $m$-representation 
with the use of modern software (in fact,
the FFTW library [46] is used), while all nonlinear
operations are local in the $\vartheta$-representation. As the
main dynamical variables, we take $\alpha(t)$, $\rho_m(t)$, 
and $D_m(t)$ with $0\le m < M$ (for negative $m$, we use the relations 
$\rho_{-m}=\bar\rho_m$ and $D_{-m}=\bar D_m$). To guarantee the stability 
of the numerical scheme at large $m$, after each
fourth-order Runge-Kutta procedure, we keep only
the spectral components with $|m|< M_{eff}$, where 
$M_{eff}\approx (1/4)N$, $M\approx (3/8)N$, 
while $N=2^{12...19}$ is the size of the
arrays used for the fast Fourier transform. During each
numerical experiment, the number $N$ is doubled several 
times, when necessary, if small-scale wave structures 
are formed on the free surface. As a result of such
an adaptive increase in $N$, the right-hand sides of the
evolution equations can be calculated with approximately 
the same numerical error of $\delta_0< N 10^{-18}$, which
corresponds to the type {\tt complex}. Since the time
integration step decreases in this case as $\tau\sim 1/N$, the
error in calculating the position of the free surface at
$t\sim 1$ can be estimated as $\delta_s\lesssim N^2 10^{-18}$. 
In practice, the integrals of motion are conserved with an accuracy of
up to 10-12 decimal places throughout most part of
the evolution. At the final stage, the greater number
$N_{final}$ is used, the later comes the moment when the
high accuracy is lost.

At the end of this section, we should say a few
words about how to set the initial configuration of the
free surface in conformal variables if its dependence is
known, for example, in polar coordinates $(r,\chi)$ in the
form of an explicit expression $r=r(\chi)$. To this end, we
should set $\rho=0$ and take the value of $\alpha_0$ so that the
volume of the hollow domain corresponds to the function $r(\chi)$. 
Then we should formally temporarily
remove the Coriolis force from the generalized Bernoulli 
equation and modify the potential of the centrifugal force:

$$\Omega^2|Z^{(s)}|^2/2\to \Omega^2(|Z^{(s)}|^2-r^2(\chi))/2.$$
          
In addition, we should set $g=0$ and add uniform linear
damping to the Bernoulli equation by replacing $\psi_t\to(\psi_t+\gamma\psi)$
in this equation. As a result of the evolution
of such a modified system, the configuration of the
free boundary quickly relaxes to the required shape,
after which we can return to the original equations and
start the computation.

\section{Brief summary of previous results}

The first numerical examples given in [35-37] set
as their primary goal to demonstrate the applicability
of the method to the description of the dynamics of
very steep (and even overturning) wave profiles over a
nonuniform bottom, rather than to verify any theoretical 
predictions. A typical formulation of numerical
experiments was to initiate an initial solitary wave over a
relatively deep bottom (or create a small group of
waves due to the movement of the bottom, thereby
simulating the tsunami phenomenon), and then
observe the evolution of this wave when it runs up onto
a shallower region. As expected from everyday experience, 
when propagating into shallow water, the wave
increases its amplitude and overturns. In some simulations, 
the process is accompanied by the formation
of blocked waves in those places where a large-scale
current passes from smaller to larger depths.

A more nontrivial phenomenon in the interaction
of waves with nonuniform bottom is the so-called
Bragg (or gap) solitons. If, in a spatially one-dimensional 
wave system (such as the free boundary over a
two-dimensional flow), waves interact with a periodic
structure, then the frequency spectrum of linear waves
is divided into zones between which gaps appear. If we
take into account nonlinearity, then, in some cases,
long-lived localized structures in the form of envelope
solitons for standing waves may arise. In this case, the
soliton frequency lies within the gap. As applied to the
waves over periodic bottom profiles (the bottom has
period $\Lambda$ along $x$, and the modulated standing wave
has length $2\Lambda$), such coherent structures were considered 
by the author in [38,39]. The theoretical predictions 
based on previously known solutions of approximate 
model equations were successfully confirmed by
direct numerical simulation of the exact equations
within the method discussed here. The Bragg solitons
given at the initial moment by an analytical formula
did not disappear sometimes during hundreds of wave
periods, although, of course, effects unaccounted for
by the weakly nonlinear theory made themselves felt in
the case of large amplitudes. On the whole, this phenomenon 
looks rather curious and paradoxical: a
localized standing wave on the water surface does not
split in time into two groups of waves running left and
right, although it would seem that the surface is free
and nothing prevents the wave from splitting. It is very
important that the most favorable regime for observing 
such gap solitons is obtained with very strong bottom 
deviations (in fact, this is a ``comb'' of narrow and
high barriers), which can hardly be equally adequately
and easily processed by other methods.

Another example of paradoxical nonlinear dynamics 
is the famous Fermi-Pasta-Ulam recurrence,
when the initial state of a wave system in the form of a
single excited spatial harmonic is first transformed by
nonlinearity into a set of solitons, and then, after a
long period of nontrivial interactions between them,
the wave energy is again almost completely concentrated 
in the initial harmonic. As applied to waves in a
finite pool, this phenomenon was first modeled in [40,41]. 
Although the bottom of the pool in this case is
assumed to be horizontal, all other significant ingredients 
of the method work to the full extent. For example,
without taking into account the time dependence
of the conformal modulus, high-precision calculations
would be impossible. In these numerical experiments, 
we took, as the initial condition, a stationary
fluid with vertical deviation of the free boundary in the
form of a half cosine wave with a maximum at one end
of the pool (at $x=0$) and a minimum at the other end
(at $x=L/2$). Further, the wave system evolved according 
to the scenario typical of the FPU phenomenon:
several slow oscillations in the standing wave regime,
and then the formation of shorter coherent structures,
solitons, and their nontrivial long-term dynamics
during which approximately standing waves with multiple 
wave numbers appeared for a short time. Then it
was as if time had been reversed, and a more ordered
state was formed in the form of the same initial cosine
wave from a less ordered state. After several standing
oscillations, the process was approximately repeated.
Depending on the relative length of the pool $L/2h$ and
the initial amplitude of the cosine, $A_0/h$, the number
of interacting solitons and the recurrence period were
significantly different. Aproximately, the recurrence
time is fitted by the formula
$$T_{\rm FPU}(g/h)^{1/2}\approx C(L/h)^2(h/A_0)^{1/2},$$
where $C=0.148+0.096(A_0/h)$. The accuracy of
recurrence to an (almost) stationary state was surprisingly 
high for not too long pools with $L/h\sim 60-100$
and not too large amplitudes of $A_0/h\sim 0.12-0.15$.

In addition, in [41], the present author took initial
states in the form of several solitons and simulated
extreme waves over various bottom profiles, including
long-wave and short-wave irregularities. In the case of
the initial state consisting of several separated solitons
over a horizontal bottom, the quality of recurrence was
generally much worse than that in the case of the
cosine wave. As for the extreme waves, the flat bottom
and the corresponding quasi-integrable dynamics
regime did not promote the emergence of extreme
waves. If the approximate integrability was violated by
the bottom irregularities, then the system passed to the
state of a random wave field, in which quasisolitonic
coherent structures of different amplitudes were present, 
some of which being much stronger than the initial 
solitons. The collision of the strongest counterpropagating 
solitons gave rise to extreme waves. The highest 
waves were observed over a smooth bottom profile,
while for relatively short-correlated bottom irregularities, 
the extreme waves were lower but had sharper
crests. Similar effects were observed both for waves
between two vertical walls and for waves with periodic
boundary conditions without walls.

The subject of anomalous waves was also touched
upon in [42], but in a slightly different setting. Analytical 
estimates for deep-water waves in the presence of
a large-scale inhomogeneous current have shown that
an opposing increasing current renders the wave more
modulationally unstable and thereby contributes to
the formation of anomalous waves. To observe this
effect in a direct numerical experiment, the author
included potential stream (an analog of $\Gamma$) over a nonuniform 
bottom profile and modeled relatively short
traveling waves. The direct interaction of such waves
with the bottom is negligible, but the presence of a
stream, slow over deep bottom and faster over shallow
bottom, was important. At the initial instant of time, a
rather long modulated wave packet was launched in
the region with a slow current, which then propagated
upstream. Upon reaching a strong opposing current,
modulational instability developed, and anomalous
waves were formed in accordance with the theory. It
was also noted that if a quasi-random sequence of
wave groups, rather than a weakly modulated long
wave packet, reaches the opposing current, then much
higher rogue waves can arise than those predicted by
the formula based on the solution of the nonlinear
Schr\"odinger equation in the form of the so-called
Akhmediev breather. The reason for the appearance of
higher anomalous waves lies in the attractive interaction 
between quasisolitonic coherent structures into
which typical wave groups turn upon reaching a fast
opposing current, while the Akhmediev breather and
the formula based on it do not take into account possible 
processes of fusion of quasisolitons.

\begin{figure}
\begin{center}
\epsfig{file=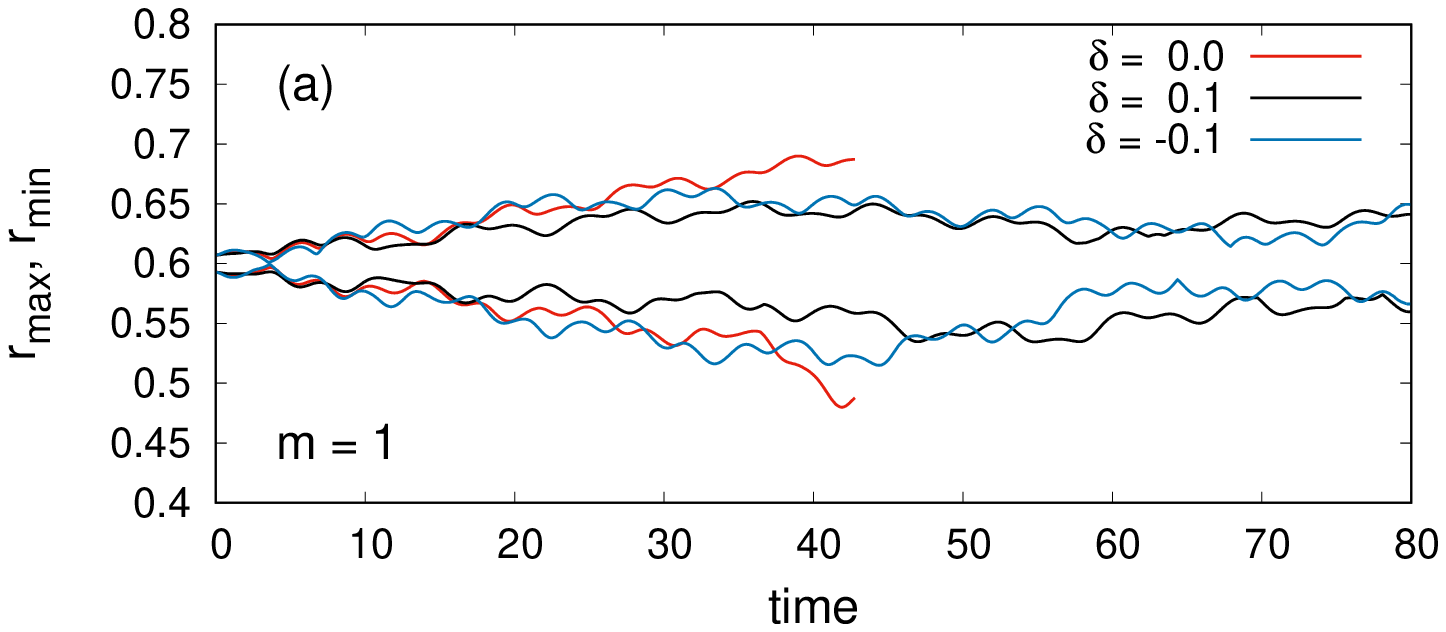, width=82mm}\\
\epsfig{file=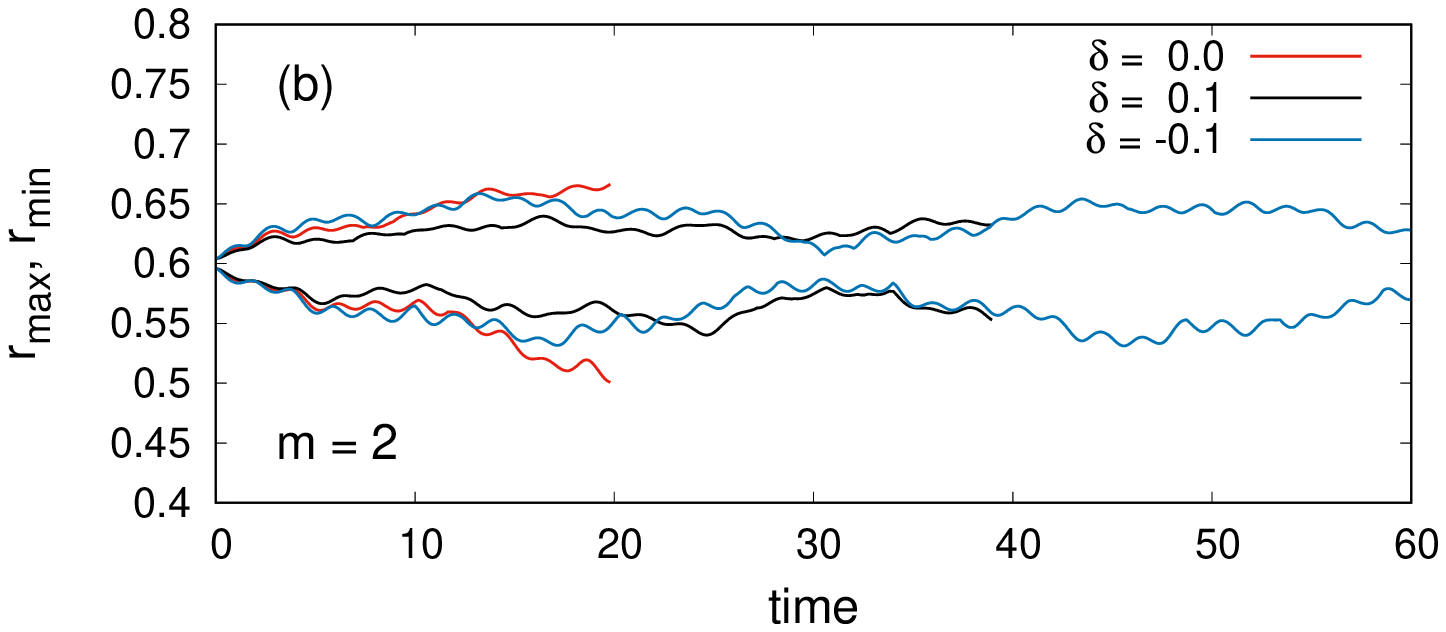, width=82mm}\\
\epsfig{file=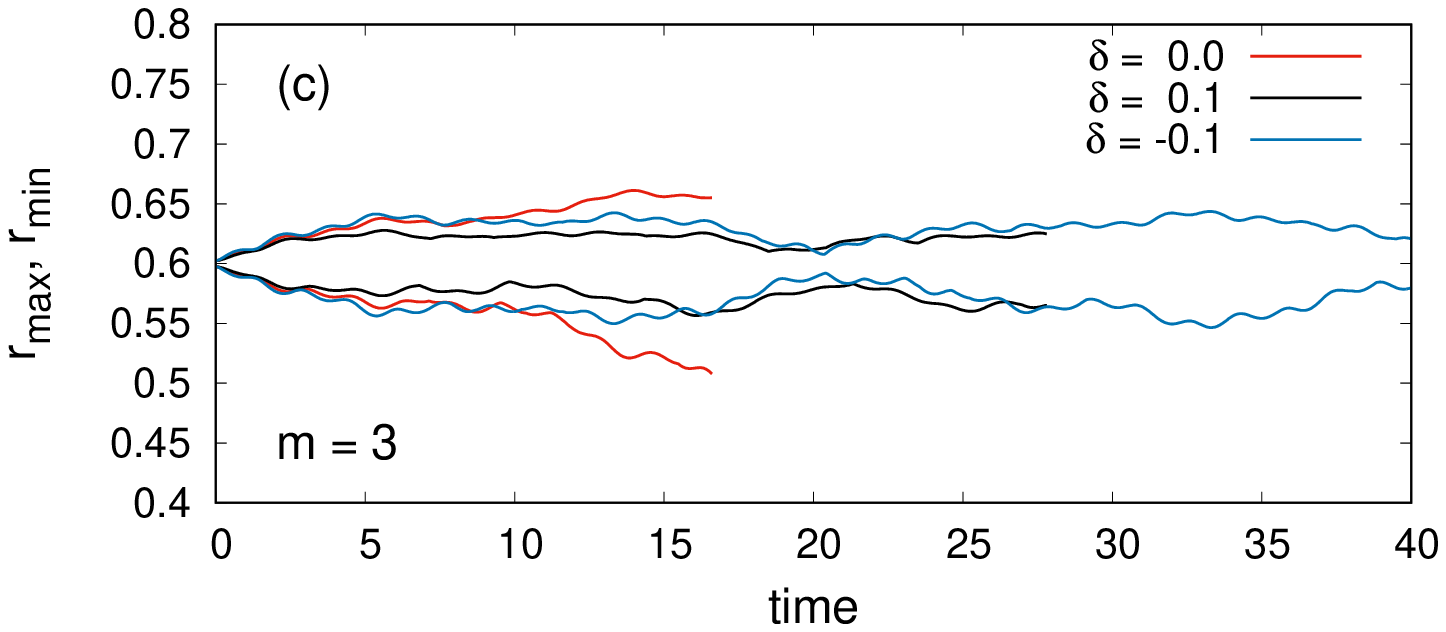, width=82mm}\\
\epsfig{file=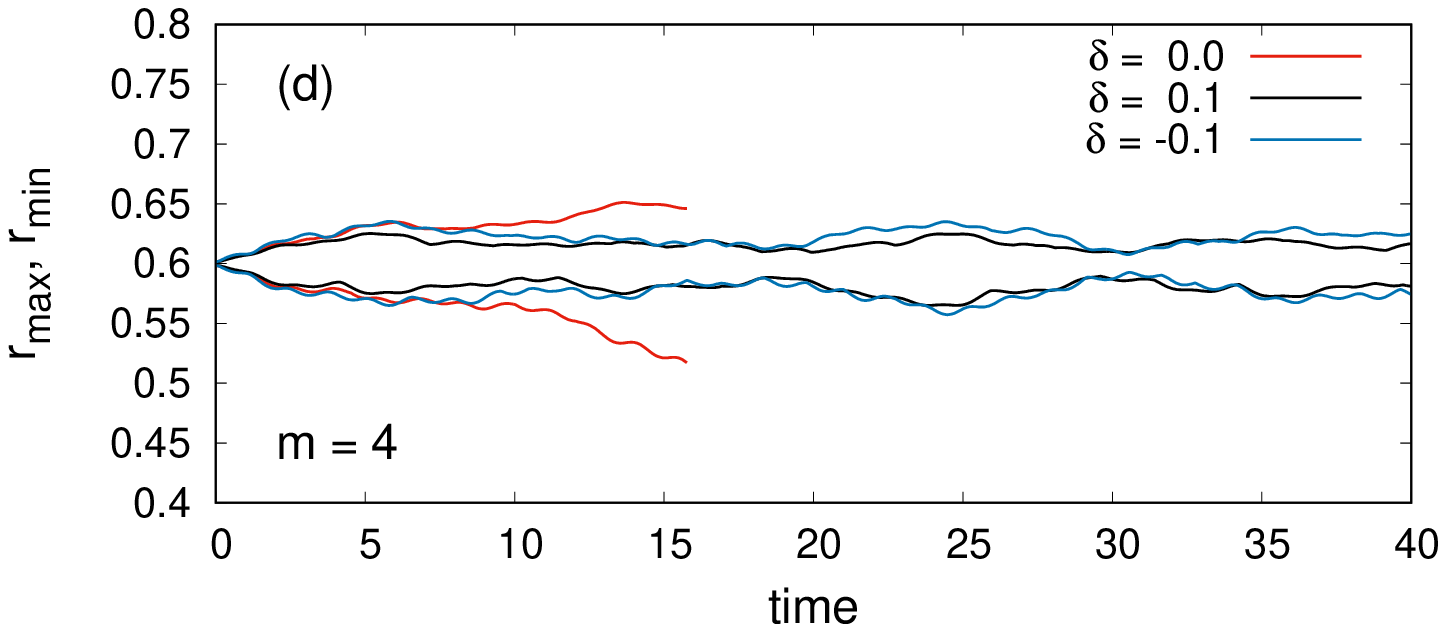, width=82mm}\\
\end{center}
\caption{Time dependence of the maximum
and minimum values of the radial coordinate of the free
surface in slightly noncircular centrifuges for different azimuthal 
numbers $m$ of the bottom deviations and different
detunings $\delta$ of the angular velocity of rotation from the
corresponding resonant values.}
\label{r_max_min_m1234} 
\end{figure}
\begin{figure}
\begin{center}
\epsfig{file=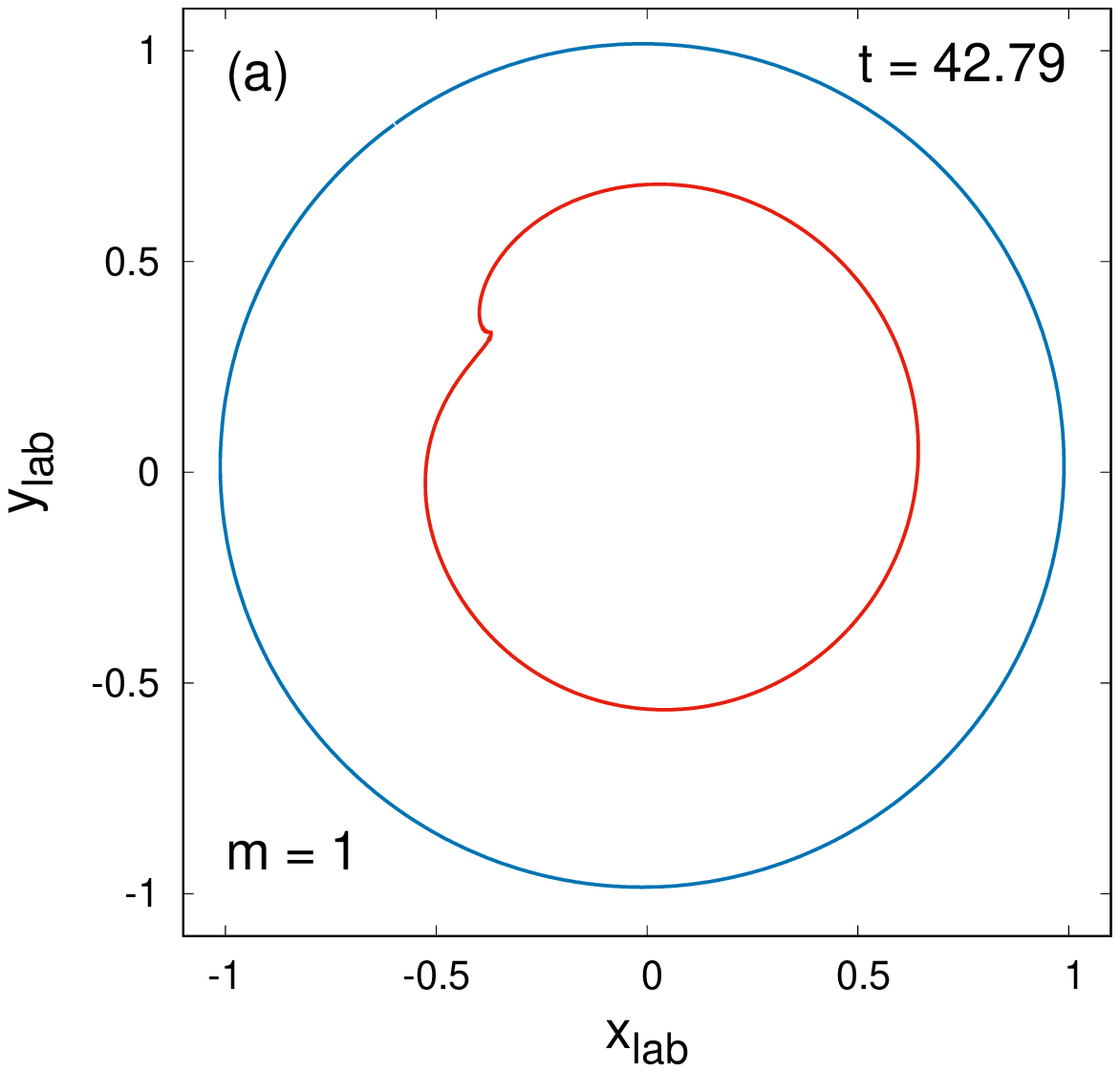, width=42mm}
\epsfig{file=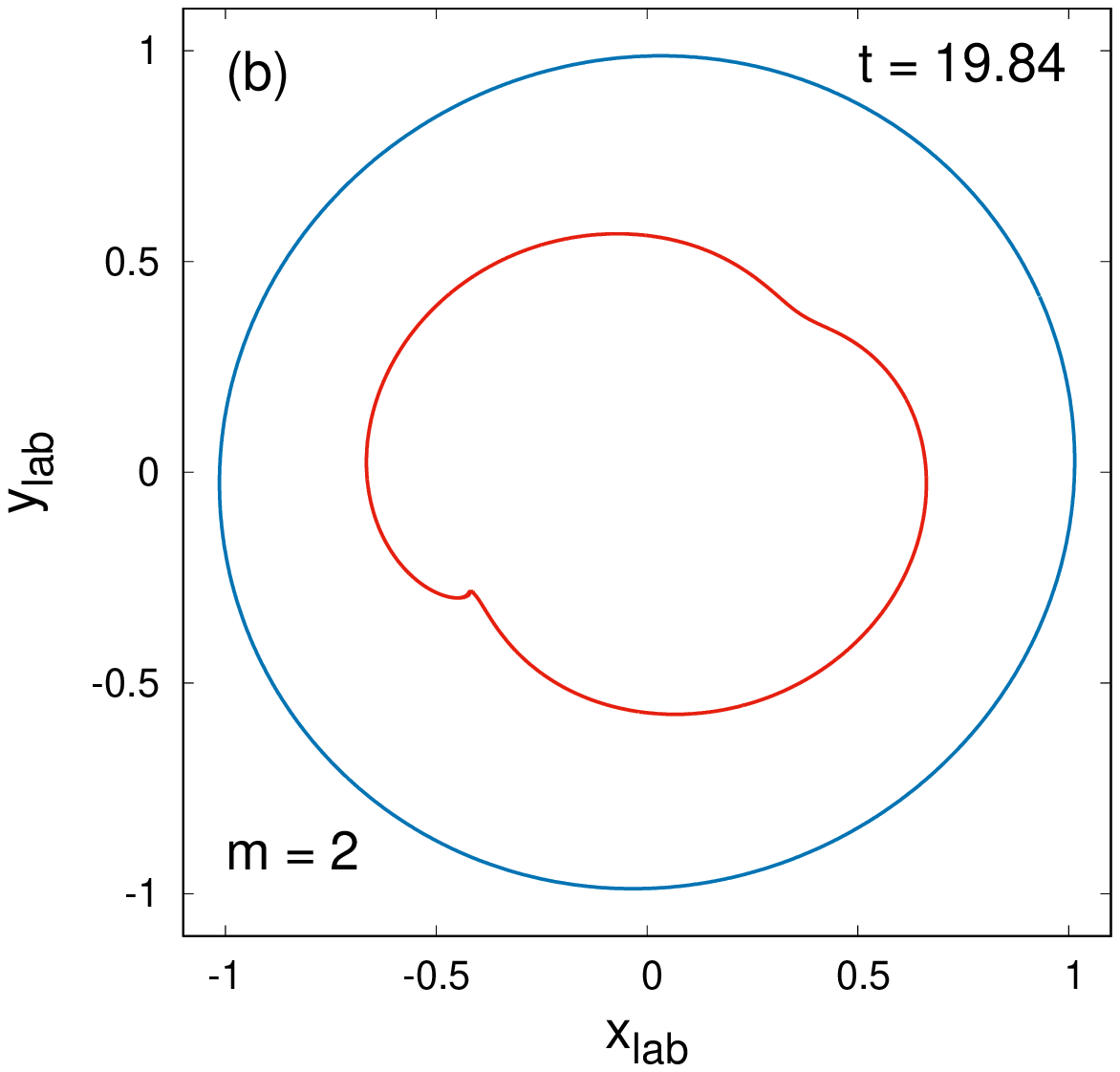, width=42mm}\\
\epsfig{file=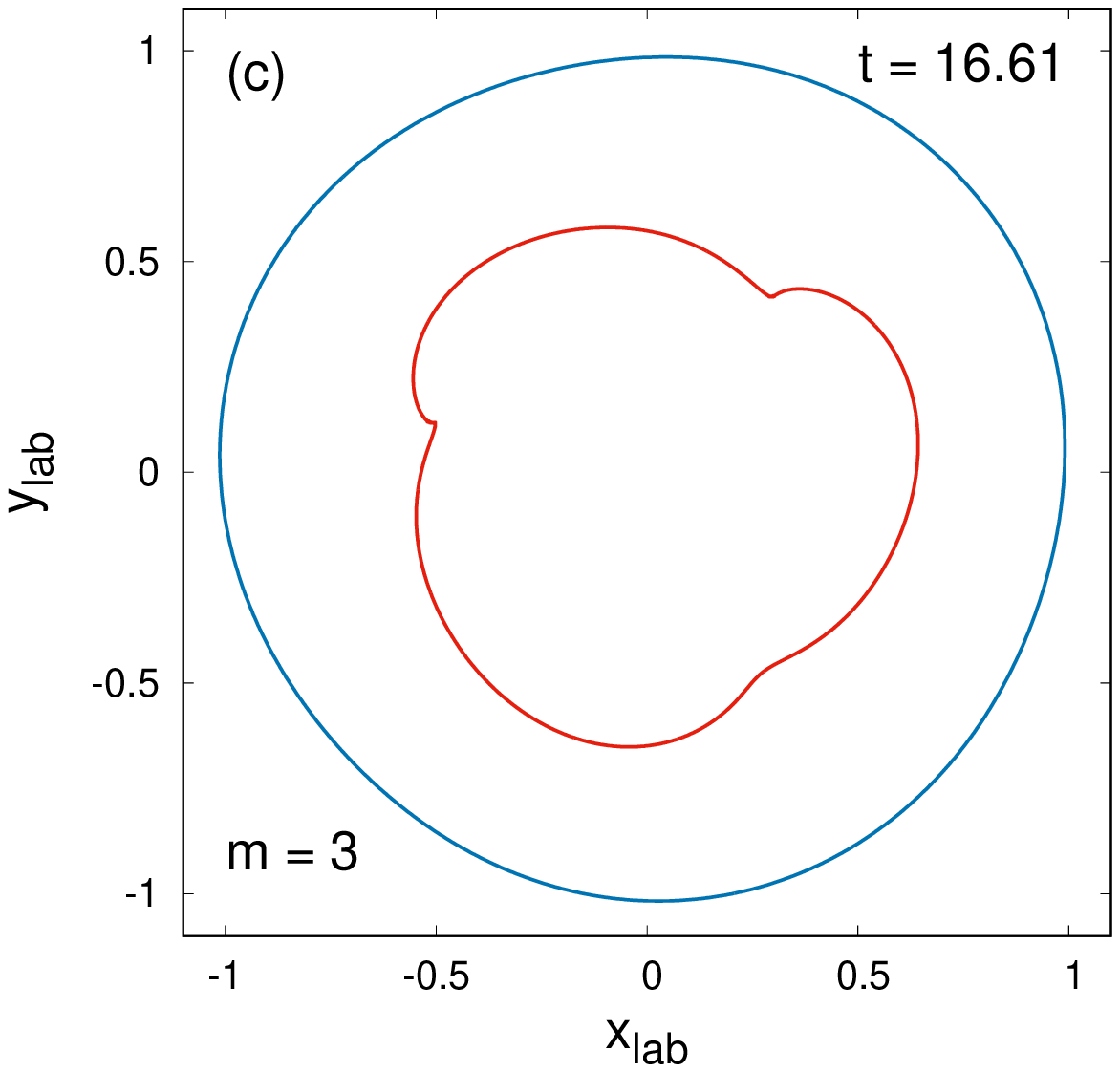, width=42mm}
\epsfig{file=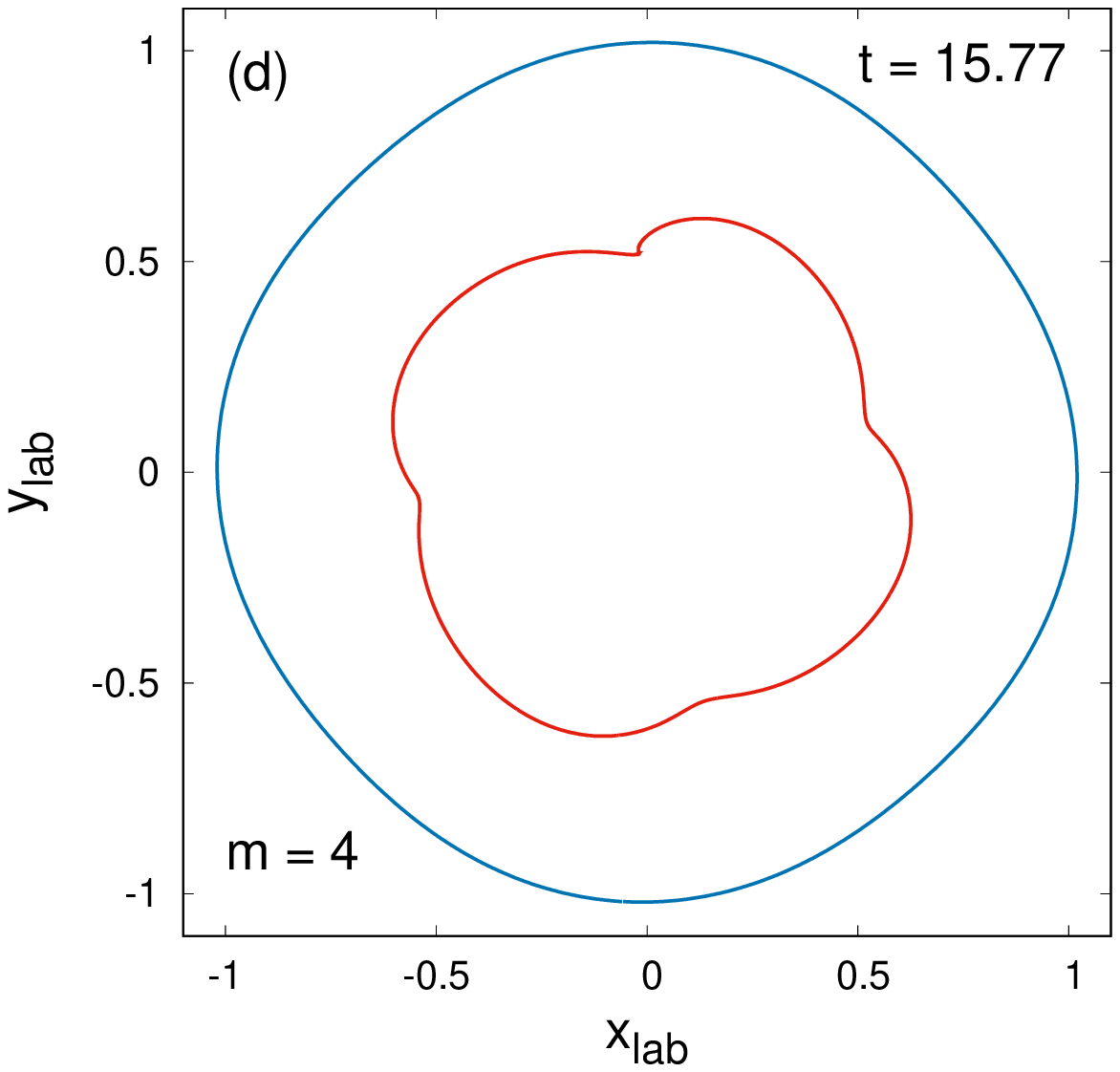, width=42mm}\\
\end{center}
\caption{Formation of singularities on the 
free surface for $\delta=0$ and different $m$.}
\label{waves_m1234} 
\end{figure}

\section{New application: waves in centrifuges}

Before proceeding to new numerical results, we
consider linearized equations of motion for surface
waves in a partially filled, perfectly circular centrifuge
(of unit radius and unit angular velocity) under the
condition of relatively small gravity force ($g/\Omega^2R \ll 1$
in dimensional variables). We will use polar coordinates 
$r$ and $\chi$, so that $x+iy=r\exp (i\chi)$. In this case, the
unknown functions are the small deviation of the free
boundary $\eta(\chi,t)=[r(\chi,t)-c]$ from the equilibrium
radius $0<c<1$ and the boundary value $\psi(\chi,t)$ of the
velocity field potential. The generalized Bernoulli
equation and the kinematic boundary condition on
the free boundary in the main approximation are as
follows (for $\Gamma=0$):
\begin{eqnarray}
\psi_t&=&c\eta-2\hat{\mathfrak R}\psi-cg(t)\sin(\chi+t),
\label{lin_eq_psi}
\\
\eta_t&=&c^{-1} \hat{\mathfrak R} \psi_\chi,
\label{lin_eq_eta}
\end{eqnarray}
where the operator $\hat{\mathfrak R}$ (an analog of the operator 
$\hat{\sf R}_\alpha$ ) is
diagonal in the discrete Fourier representation:
\begin{equation}
{\mathfrak R}_m=i\frac{(1-c^{2m})}{(1+c^{2m})}=i\tanh(m\ln(1/c))\equiv i\sigma_m.
\end{equation}
The relation $\theta(\chi,t)= \hat{\mathfrak R}\varphi(\chi,t)$ 
for $r=c$ arose from the
fact that the fulfillment of the kinematic condition on
the container wall is ensured by equality
\begin{equation}
\phi_m=A_m(t)r^{m}e^{im\chi}+\bar A_m(t)r^{-m}e^{-im\chi}.
\end{equation}
In the case of $g=const$, a particular solution of the
inhomogeneous linear system (35), (36) has the form
\begin{eqnarray}
\psi_{\rm st}&=&\frac{cg}{(1+\sigma_1)}\cos(\chi+t),\\
\eta_{\rm st}&=&\frac{-\sigma_1g}{(1+\sigma_1)}\sin(\chi+t).
\end{eqnarray}
It corresponds to the stationary state in the laboratory
coordinate system (see [43] for details). Equations
(35) and (36) also yield an expression for the eigenfrequencies 
(the ``dispersion law''; incidentally, we note
that, in [43], the dispersion equation was derived in a
nonrotating system, and this is probably why the
answer was not brought to such a simple formula)
\begin{equation}
\omega_m=\sigma_m +\sqrt{\sigma_m^2+m\sigma_m}\,,
\end{equation}
here positive $m$ correspond to leading waves, and negative $m$,
to lagging waves (in laboratory coordinates). It
is clear that if perturbations with azimuthal number $m$
and relative rotation speed $\Delta\approx\omega_m/m$ are introduced
into the system, then one should observe a resonant
growth of the corresponding harmonic. Such perturbations 
are most easily implemented by slightly distorting 
the shape of the centrifuge and forcing it to
rotate with frequency of about $1+\Delta$, rather than with
frequency $\Omega=1$. In the numerical examples shown in
Figs. 1 and 2, such a regime is provided by the function
\begin{equation}
Z(\zeta,t)=\exp[i(-\omega_{-m}/m+\delta) t]e^{i\zeta}[1+\epsilon e^{im\zeta}]
\end{equation}
with $\epsilon=0.02$  for $m=1,2,3,4$. The detuning from the
resonance was fixed by the parameter $\delta$. As the initial
conditions, we took the functions $\rho=0$ and $\psi=0$, and
the initial value $\alpha_0=-\ln(0.6)$ gave an approximately
circular shape of the free boundary with an average
radius of $c\approx 0.6$. In this series of numerical experiments, 
we used a dimensionless gravity field of $g=0.02$.

Figures 1 and 2 show that there really was a resonant growth 
of waves accompanied by an enhancement 
of their nonlinearity and ending with the formation 
of a singularity on the free surface. In real conditions, 
this would mean the beginning of the transition
of the flow to a three-dimensional turbulent regime.

\begin{figure}
\begin{center}
\epsfig{file=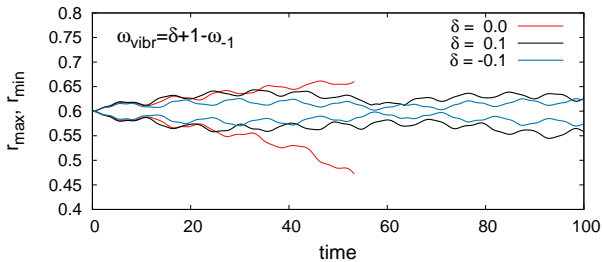, width=82mm}
\end{center}
\caption{The maximum and minimum values
of the radial coordinate of the free surface as a function of
time for vertical vibration of the axis of a circular centrifuge
with effective gravity force $g(t)=0.02-0.01 \cos(\omega_{\rm vibr}t)$.
}
\label{r_max_min_vibr} 
\end{figure}
\begin{figure}
\begin{center}
\epsfig{file=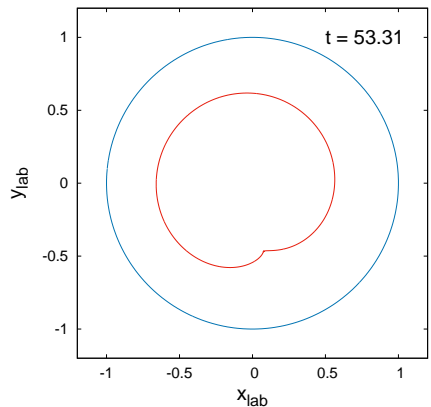, width=60mm}
\end{center}
\caption{Formation of a singularity on the
free surface in a circular centrifuge with a vertically vibrating 
axis for a resonant value of the vibration frequency.
}
\label{wave_vibr} 
\end{figure}

Another possibility of introducing resonant perturbations 
is given by the vertical vibrations of the axis of
rotation, which lead to the dependence
$$g(t)=g_0+A\cos(\omega_{\rm vibr} t).$$
Such perturbations act most effectively on the harmonics 
with $m=\pm 1$, even in the case of a perfectly circular 
container, as follows from the approximate equations 
(35) and (36). The resonance frequencies are
given by the formula $\omega_{\rm vibr}=1\pm\omega_{\pm 1}$. 
The corresponding examples are illustrated in Figs. 3 and 4.

\begin{figure}
\begin{center}
\epsfig{file=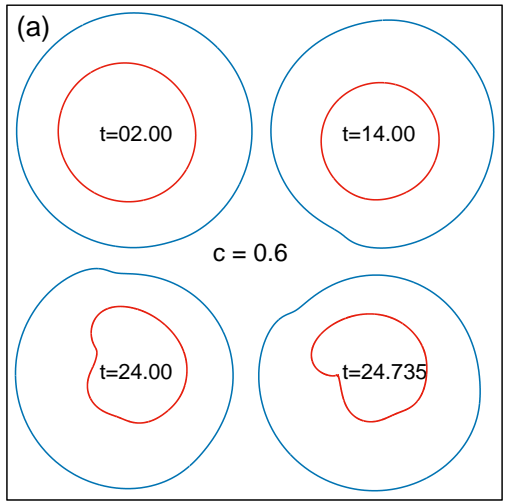, width=74mm}\\
\epsfig{file=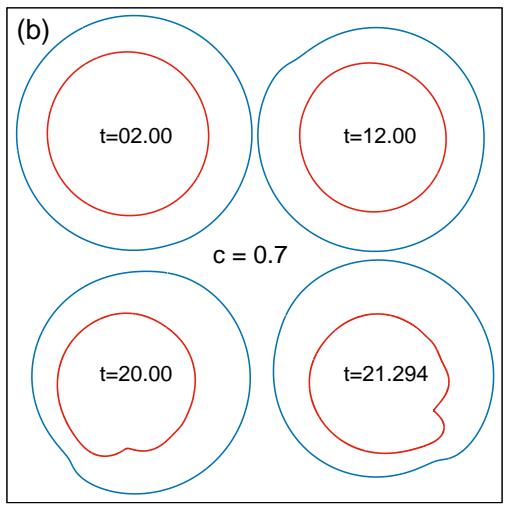, width=74mm}\\
\epsfig{file=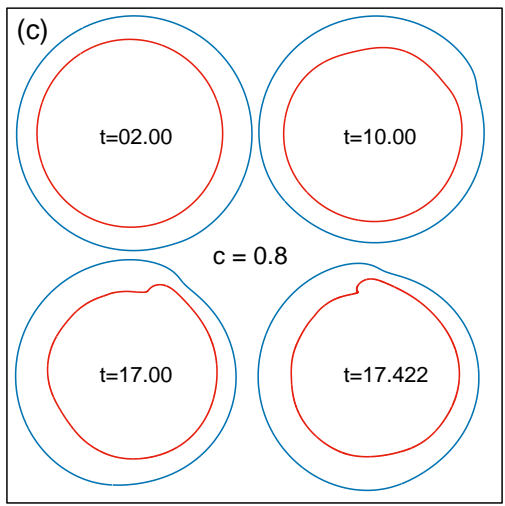, width=74mm}
\end{center}
\caption{Formation of a singularity on
the free surface in a centrifuge deformed according to
formula (43) with $m=1$, $C_1=0.9$, and $C_2=0.7$ 
for three values of the initial radius.
}
\label{wave_deform1} 
\end{figure}
\begin{figure}
\begin{center}
\epsfig{file=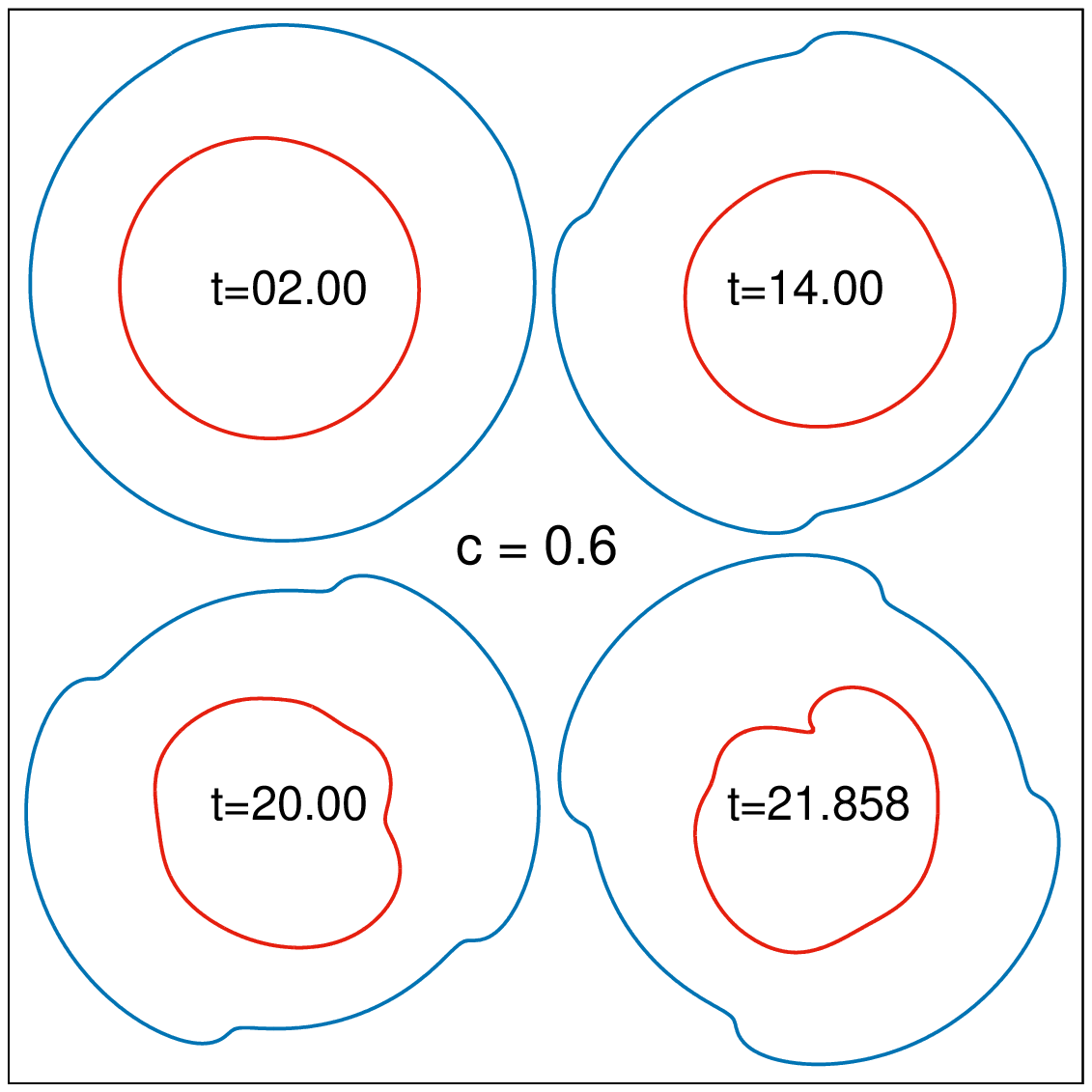, width=78mm}
\end{center}
\caption{Formation of a singularity on the
free surface in a centrifuge deformed according to formula
(43) with $m=2$, $C_1=0.9$, and $C_2=0.9$.
}
\label{wave_deform2} 
\end{figure}
\begin{figure}
\begin{center}
\epsfig{file=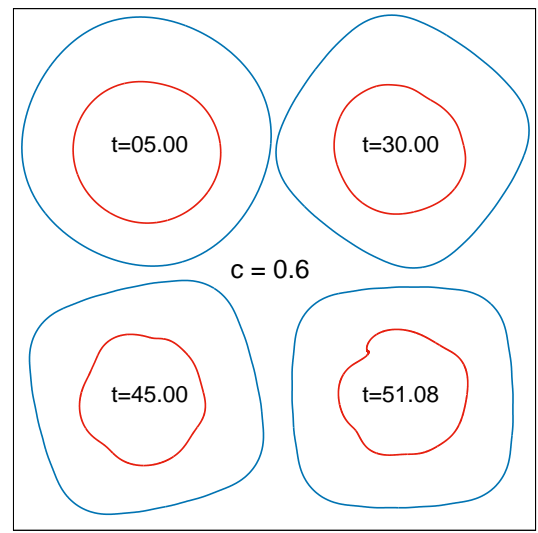, width=78mm}
\end{center}
\caption{Formation of a singularity on the
free surface in a centrifuge deformed from a circle to a
square according to formula (46).
}
\label{wave_deform_square} 
\end{figure}

To demonstrate the full potential of the method, in
Figs. 5 and 6 we show examples of the evolution of the
free surface in a deformable centrifuge with a cross-sectional 
area varying (decreasing) with time. In these
experiments, $g=0.1$, and the shape of the container
corresponded to the analytic function
\begin{equation}
Z(\zeta,t)=\exp\Big[i\zeta +id(t)\ln\Big(\frac{1+C_1e^{im\zeta}}
{1-C_2e^{im\zeta}}\Big)-0.5\pi d(t)\Big],
\label{deform_m}
\end{equation}
where the measure of deviation from circular shape
was given by the expression $d(t)=0.1[1-\exp(-0.04 t)]$.
In Fig. 5, the parameters are $m=1$, $C_1=0.9$, and $C_2=0.7$. 
In Fig. 6, we took $m=2$, $C_1=0.9$, and $C_2=0.9$.
At the initial time, this was a circularly symmetric configuration 
with the radius of the free surface $c$. Then,
part of the bottom of the container was seemingly
raised along the radius, thus forming $m$ regions with
``smaller depth.'' In this case, the decrease in the container 
area was accompanied by the appearance of an
additional counterclockwise circular potential stream
with parameter $\Gamma(t)$. Singularities (overturning angles)
were formed on the crest of a growing wave near the
point where the stream passed from smaller to larger
depth.

Finally, Fig. 7 shows the deformation of the initially 
circular container to a shape intermediate
between a circle and a square (for $g=0.1$, $c=0.6$). In
this numerical experiment, we used the expansion of
the corresponding elliptic integral up to the fourth
order,
\begin{eqnarray}
&&\int_0^z\frac{dz}{\sqrt{1-z^4}}\approx zE_4(z^4),\\
&&E_4(\mu)=\Big(1+\frac{\mu}{10}+\frac{\mu^2}{24}+\frac{5\mu^3}{16\cdot 13}
+\frac{35\mu^4}{128\cdot 17}\Big),
\end{eqnarray}
to construct the function
\begin{equation}
Z(\zeta,t)=[1-0.1\tilde d(t)]e^{i\zeta}E_4\big(\tilde d(t)\exp(4i\zeta)\big),
\label{deform_sq}
\end{equation}
where $\tilde d(t)=0.8[1-\exp(-0.04 t)]$. In this case, a singularity 
was also formed. However, if, instead of the
factor $[1-0.1\tilde d(t)]$, we took the factor $[1-0.2\tilde d(t)]$,
which resulted in a more significant decrease in the
container area and the hollow domain in it after deformation, 
then (for $c=0.6$) the waves on the free surface
remained smooth for a long time and did not show a
tendency to form sharp overturning crests (these
results are not shown). There is yet no complete
understanding of the reasons for such behavior.

\section{Conclusions}

Thus, the main result of this work is the development 
and application of the method of composite
conformal mapping to describe the dynamics of waves
on the free surface of an ideal incompressible fluid in
partially filled flat centrifuges of complex shape. Just
as in previous applications, the method has demonstrated 
high accuracy and efficiency. True, so far we
cannot say that all questions have been exhausted. For
example, it is not yet clear how to modify the method
in the case when a relatively small hollow domain (a
hollow vortex for $\Gamma\neq 0$) moves far from the origin
during its dynamics so that the origin turns out to be
filled with a fluid. Obviously, a conformal mapping of
the type $Z(\zeta)=\exp(i\zeta)$, which has a singularity at  $Z=0$
(or at any other previously fixed point), fails to work
in this case.

However, even in the existing version, the method
can solve many interesting problems. It is quite clear
that the examples presented here are far from exhausting 
the entire variety of possible wave structures and
their dynamics in rotating systems with nontrivial
geometry.

\end{document}